\documentclass[singleblind]{ecai}  

\usepackage{graphicx}
\usepackage{latexsym}
\usepackage{booktabs}
\usepackage[inline]{enumitem}
\usepackage[many]{tcolorbox}

\ecaisubmission      

\paperid{1538}        

\newtcolorbox[auto counter]{recommendation}[2][]{fonttitle=\bfseries,
title=Recommendation~\thetcbcounter #1, label=#2}

\begin{document}

\begin{frontmatter}

\title{AI Research is not Magic, it has to be Reproducible and Responsible: Challenges in the AI field from the Perspective of its PhD Students}

\author[A]{\fnms{Andrea}~\snm{Hrckova}\orcid{0000-0001-9312-6451}\thanks{Corresponding Author. Email: andrea.hrckova@kinit.sk.}}
\author[B]{\fnms{Jennifer}~\snm{Renoux}\orcid{0000-0002-2385-9470} }
\author[C]{\fnms{Rafael }~\snm{Tolosana Calasanz}\orcid{0000-0003-3057-6273}} 
\author[A]{\fnms{Daniela}~\snm{Chuda}\orcid{0000-0002-3873-9308}}
\author[A]{\fnms{Martin}~\snm{Tamajka}\orcid{0000-0002-0107-6459}}
\author[A]{\fnms{Jakub}~\snm{Simko}\orcid{0000-0003-0239-4237}}

\address[A]{Kempelen Institute of Intelligent Technologies}
\address[B]{Orebro University}
\address[C]{University of Zaragoza}

\begin{abstract}
With the goal of uncovering the challenges faced by European AI students during their research endeavors, we surveyed 28 AI doctoral candidates from 13 European countries. The outcomes underscore challenges in three key areas: (1) the findability and quality of AI resources such as datasets, models, and experiments; (2) the difficulties in replicating the experiments in AI papers; (3) and the lack of trustworthiness and interdisciplinarity. From our findings, it appears that although early stage AI researchers generally tend to share their AI resources, they lack motivation or knowledge to engage more in dataset and code preparation and curation, and ethical assessments, and are not used to cooperate with well-versed experts in application domains. Furthermore, we examine \emph{existing practices in data governance and reproducibility} both in computer science and in artificial intelligence. For instance, only a minority of venues actively promote reproducibility initiatives such as reproducibility evaluations.

Critically, there is \emph{need for immediate adoption of responsible and reproducible AI research practices}, crucial for society at large, and essential for the AI research community in particular. This paper proposes a combination of social and technical recommendations to overcome the identified challenges. Socially, we propose the general adoption of reproducibility initiatives in AI conferences and journals, as well as improved interdisciplinary collaboration, especially in data governance practices. On the technical front, we call for enhanced tools to better support versioning control of datasets and code, and a computing infrastructure that facilitates the sharing and discovery of AI resources, as well as the sharing, execution, and verification of experiments.




\end{abstract}

\end{frontmatter}

\section{Introduction}

AI research is facing a reproducibility crisis~\cite{cockburn2020threats, vollmer_machine_2020}. Worse, AI systems and methods are increasingly being used to perform research in other fields, and concerns arise that it may lead to another major crisis in science in general, as scientists may use non-reproducible and non-responsible AI systems in an ill-informed way~\cite{ball2023is}. 
In this work, we use the term \textit{reproducibility} as \textit{obtaining consistent results using the same input data; computational steps, methods and code; and analysis conditions}~\cite{national2019understanding}.

Multiple reasons for the crisis have been investigated, such as data quality and the reproducibility of code and models. For instance, Paullada et al.~\cite{paullada_data_2021} argue that data challenges are often overlooked in machine learning and criticize the insufficiently careful practices of data annotation and documentation. Furthermore, research by Wang et al.~\cite{wang_devil_2018} and Reale et al.~\cite{reale_analysis_2016} point out the impact of dataset quality on model performance. 

In an attempt to make AI research more reproducible and responsible, there are occasional examples of universities that integrate reproducibility into AI ethics education~\cite{Lucic_Bleeker_Jullien_Bhargav_deRijke_2022}. Moreover, many high-impact venues are imposing ``Reproducibility Checklist''\footnote{\url{https://aaai.org/aaai-conference/reproducibility-checklist/}}, ``Reproducibility Guidelines''\footnote{\url{https://ijcai24.org/reproducibility/}} or ``Artifact Badges''\footnote{\url{https://www.acm.org/publications/policies/artifact-review-and-badging-current}}. 
Such initiatives are mostly one-offs, limited to when the paper is submitted, and there is no monitoring of the results and code in the long term. In addition, they usually focus on availability and proper documentation of data, either self-reported or evaluated by the reviewer, though without any incentive to actually attempt to make the research described in the research papers more reproducible. An exception to this are ``Reproducibility challenges'' introduced recently in some Machine Learning conferences~\cite{pineau2021improving}, where independent teams of researchers attempt to verify the empirical claims of a published paper and produce a ``reproducibility report''. Nonetheless, the long-term impact of these initiatives on the quality of the work is still an open question~\cite{pineau2021improving}.

To better understand the state of AI research reproducibility and its challenges, \textbf{this paper presents the findings of an in-depth, qualitative survey of 28 European AI PhD students}. The survey investigates the practices and hurdles encountered by the AI research community in general, and observes reproducibility-related issues emerging most prominently. The PhD candidates (more broadly: early-career researchers) are often the ones, who undertake the labor of reproducing previously published results and are thus most exposed to reproducibility hurdles. It is thus important to listen to what they have to say about these issues. This paper sheds light on numerous consequences stemming from the lack of reproducibility and responsible AI research practices within the AI community. Following on the survey outcomes, \textbf{we propose recommendations aimed at mitigating some of the issues identified}.

Our research is interdisciplinary, drawing on qualitative methods from the social sciences, primarily from \emph{information science}. Existing studies on information interaction (or \emph{information seeking behavior}, as called in this field) have brought insight into the needs and practices of academic scientists~\cite{hemminger_information_2007}, doctoral students~\cite{steinerova_methodological_2013}, and computer scientists~\cite{tucci_assessing_2011, athukorala_information-seeking_2013}. However, a gap persists in this research among AI doctoral students, which we address in this study.

The main contributions of this paper are:
\begin{enumerate}
\item We explore and express the issues encountered by European doctoral students in AI field, many of which are particularly related to reproducible and responsible AI. We identify the main sources of difficulties, which include, but are not limited to, the quality of AI resources (including datasets, code, and models).
\item We formulate recommendations to address these issues, highlighting, among others, 1.) the need for adoption of reproducibility practices on every level, particularly in AI journals and conferences, 2.) the need for improved interdisciplinarity in AI, especially greater participation of domain experts, data and information specialists, ethicists, and legal professionals (notably in data governance practices). 
\end {enumerate}

The remainder of this paper is organized as follows. Section \ref{sec:methodology} describes the methodology we used for our study, and Section \ref{sec:findings} presents our findings. Section \ref{sec:recommendations} describes the recommendations we drew from our analysis. Finally, Section \ref{sec:conclusion} discusses our work in the field of AI and suggests directions for future research. Note that throughout the paper, we use the term ``AI resource'' to refer to any artifact created as part of AI research and development, namely datasets, code and software systems, and models.

\section{Methodology}
\label{sec:methodology}
When designing the interview questions for respondents, we focused on the human-information interaction processes, a very common approach in information science research~\cite{case_looking_2016}. The questions covered the entire doctoral research process, from searching for information sources to publishing findings. We also included questions on problematic aspects of the work, which are common in the software requirements specification. In addition to that, the questions were selected based on the experience of one of the authors as a coordinator of doctoral studies. The full list of interview questions can be found in Annex A.

A total of 48 questions of the survey examined \emph{practices and difficulties} in five areas:
 \begin{enumerate}
\item Working with data and employing AI/ML methods within a particular focus and domain of AI.
\item Doing AI experiments, using infrastructure and tools for data management, programming, modelling, prototyping, benchmarking, and using methods for responsible AI and user research  
\item Information and data retrieval.
\item Organization of AI resources.
\item Publication and dissemination of research results.
\end{enumerate}

 We surveyed 28 doctoral students representing 13 different European countries\footnote{Countries, where respondents study: Slovakia, Greece, Germany, Spain, Sweden, Ireland, France, Bulgaria, Belgium, Norway, Finland, Czech republic, Ukraine}. We conducted 11 semi-structured focus group interviews with no prior hypotheses, yet with this research question: \textit{What are the biggest challenges that the doctoral students face, when doing AI research?} The time interval for the interviews was February to June 2023 due to the time constraints of the participants. Nearly all interviews were conducted online using Google Meet in English. One meeting was held face-to-face in the native language. On average, the interviews lasted 1.5 - 2 hours. One round (with two interviewees) was conducted in written form at the request of the participants.

 One focus group involved up to four participants. The participants had diverse AI focuses, spanning ten areas \footnote{AI focuses of the respondents: machine learning (encompassing neural networks, deep learning, federated learning, and accelerators), natural language processing (NLP), explainable AI, ethical AI, computer vision, recommenders, robotics, multimodal processing, and human activity recognition} and four additional domains. \footnote{Domains of research of the respondents: AI security (cyber attacks), medical data processing, AI in business and AI in energy and green environment (specifically, time-series analysis).} 16 respondents were in an advanced stage of their doctoral studies, while 12 respondents were early-stage doctoral students.
 The semi-structured interview design enabled us to delve more deeply into the areas where the actual respondents demonstrated some proficiency. 

The obtained data was analyzed using the manual content analysis method. Content analysis is a \textit{ research technique to make replicable and valid inferences from texts to the contexts of their use}~\cite{krippendorff2018content}. These inferences are derived from the text via a process known as coding \footnote{Coding in content analysis is the process of labeling and organizing the qualitative data to identify different themes and the relationships between them}. A total of 32 codes were created inductively for problems in the field of AI, 5 codes were not saturated (only one respondent had the respective problem), and 9 codes were merged with the final codes after discussions. Finally, 20 categories were selected for visualizations using mindmaps. The urgency of the findings is differentiated by colors in the mindmap visualizations, ranging from the darkest red, which indicates that more than four rounds of focus groups agreed on such a problem, through the pink, which denotes the agreement of three rounds, to the lightest pink that indicates that two rounds of focus groups agreed on that point. Furthermore, we offer the precise count of groups that agreed on the respective issues in each category.

\section{Findings}
\label{sec:findings}
Most of the challenges that we identified concern the quality of available AI resources, including datasets, code, and models; therefore, we grouped them into these three categories. These identified difficulties consumed most of the time of these researchers and significantly prevented them from being able to replicate the results in papers. As respondent P1 stated:
\begin{quote}
\emph{``Sometimes you don´t understand how they gained such results in the paper and how to replicate it.``}
\end{quote}

Yet, replication studies are the type of research in which these early researchers are most involved. Sections \ref{sec:quality-datasets}, \ref{sec:quality-code}, and \ref{sec:quality-models} explore these issues in more detail. 

However, it is worth noting that we also uncovered another set of challenges that doctoral students encounter during AI research, particularly concerning the absence of interdisciplinary collaboration, issues with information retrieval, concerns regarding the lack of human involvement, and a lack of motivation to share research findings with the public. These challenges were named "Other challenges regarding AI research process" and are explored in Section \ref{sec:interdisciplinary}.

\subsection{Quality and findability of datasets}
\label{sec:quality-datasets}
\begin{figure}[t]
\centerline{\includegraphics[height=3.5in]{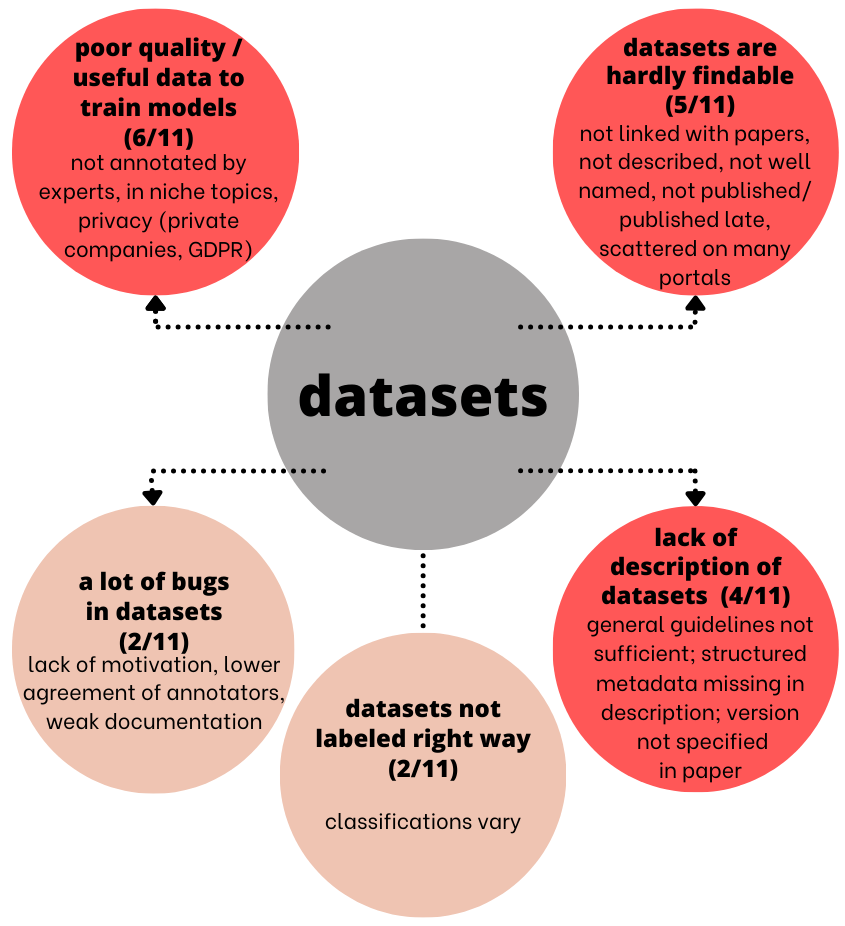}}
\caption{Dataset-related challenges of AI resources as reported by research participants. The darker the red, the more often was the problem mentioned during the focus groups. The numbers in brackets indicate the count of focus groups out of the total 11 groups that agreed on such issues.} \label{datasets}
\end{figure}

The foremost challenge frequently discussed in our interviews was the quality and utility of datasets required for the training of AI models (6 groups out of 11, see Fig.~\ref{datasets}). This challenge stems partly from the significant time investment required to curate datasets annotated by human annotators. Some respondents (6 groups out of 11) noted difficulties in accessing experts, low agreement between annotators (2 groups out of 11), or the necessity to involve themselves or the not-so-motivated students to annotate (2 groups out of 11). However, the main concern was the issue of privacy when attempting to publish a dataset (6 groups out of 11). Respondent P2 illustrates this problem: 

\begin{quote}
\emph{``There is just one public dataset of breast cancer that was properly labeled by experts with data about patients and everybody is using it... Obtaining approvals is difficult from patients.``}
\end{quote}

The challenge of accessing quality datasets extends beyond the medical domain. Several respondents encountered obstacles due to privacy restrictions imposed by their companies, preventing them from publishing papers despite having ample data from their workplace. This restriction poses a significant hurdle, as publishing without accompanying data is problematic. Our respondents conveyed skepticism regarding anonymization as a solution, citing doubts about its effectiveness and the impracticality of individually informing each affected individual. 

 Five groups agreed that locating good datasets can be a challenge, even when they do exist. Many datasets lack direct links to associated papers for various reasons, such as delayed publication or non-publication. Consequently, these datasets are often dispersed on multiple data storage platforms. Additionally, authors may not always consider the discoverability of their datasets when publishing, neglecting to provide clear and descriptive names or descriptions. Clear and comprehensive descriptions in introduction, metadata, and classification are essential to ensure that a dataset is not only easily discoverable, but also allows researchers to assess its relevance to their specific requirements. For a PhD student in AI, the process of repeatedly downloading and opening available datasets to discern their contents consumes a significant amount of time. An example of useful metadata to save their time would include specifying the version of the dataset used in the paper, given the multitude of available dataset versions. Although general guidelines can assist in describing datasets to a certain extent, respondents (for example, those from the security sector) expressed concern that they may not be directly applicable in specific domains.

Besides that, 4 groups of doctoral students out of 11 voiced their frustration regarding numerous bugs present and not documented in the available datasets, exemplified by respondent P3:
\begin{quote}
\emph{``Papers do not mention data drifts,  there is usually no information about the data preparation.``}
\end{quote}

Apparently, the issue is shifting from availability to quality of the dataset. An exception was seen in niche topics that still suffer from a shortage of data for training models.

\subsection{The quality of code}
\label{sec:quality-code}
\begin{figure}[b]
\centerline{\includegraphics[height=3in]{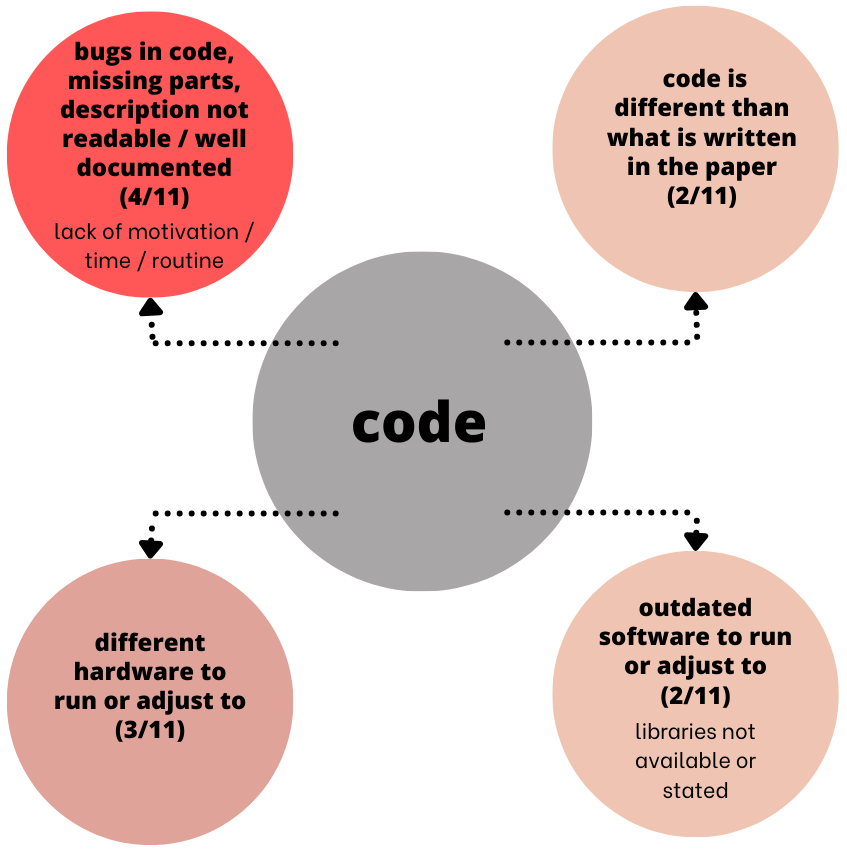}}
\caption{Code-related challenges of AI resources as reported by research participants.} \label{code}
\end{figure}

Quality issues also arise concerning the code (Fig.~\ref{code}). Early career researchers typically have a practice of publishing code (just one student mentioned the unavailable code as a problem). Our respondents were driven by inner motivations such as a sense of reciprocity or awareness of reproducibility concerns. As respondent P4 pointed out:
\begin{quote}
\emph{``If a failure is identified, it is good for science, even if it is embarrassing for an individual. ``}
\end{quote}

Nonetheless, this does not mean that the code published is clean or well documented. That is the reason why several students (4 groups out of 11) complained about serious bugs in the code or missing parts of the code. Respondent P5 highlights an additional unpleasant problem - a mismatch between the code and the paper, a concern echoed in two focus groups: 
\begin{quote}
\emph{``The biggest problem comes, if you are trying to make the paper work. Also with access to the code you get specific results. Sometimes the code is different than what is written in the paper and you have to do double work. ``}
\end{quote}

At the same time, respondents in these groups acknowledged that they lack motivation to publish meticulously documented code, as it is not a mandatory aspect of the review process and requires significant time investment, necessitating a consistent routine for continuous documentation during code writing and editing. In addition to that, these respondents emphasized that having some code available is better than having none at all.

Nevertheless, researchers would appreciate at least some information on the program versions, the required hardware, and the libraries that were used when the code was originally run. This information would save researchers a lot of time, as they quite often encounter challenges related to incorrect hardware (3 groups out of 11) or software configurations (2 groups out of 11) necessary to run and adapt the code. As noted, research papers often lack the space to address such issues.

\subsection{Benchmarking and quality of models}
\label{sec:quality-models}
\begin{figure}[b]
\centerline{\includegraphics[height=3in]{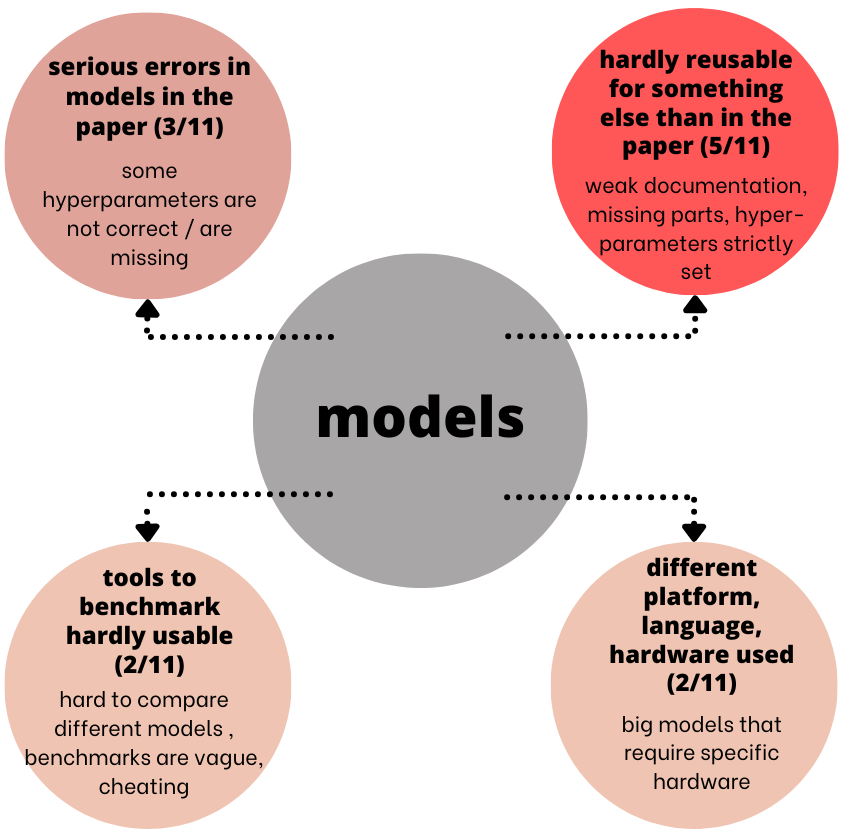}}
\caption{Model-related challenges of AI resources as reported by research participants.} \label{models}
\end{figure}

Some AI models face comparable challenges with respect to replicability and reproducibility as code (Fig.~\ref{models}). Three groups of doctoral candidates highlighted discovering significant flaws in the models presented in the papers, such as absent or inaccurate hyperparameters, along with inadequate documentation. These shortcomings, compounded by rigidly defined hyperparameters, hindered the utility of published models beyond their intended paper applications (even 5 groups of respondents out of 11 reported this problem). In addition, two institutions lacked the hardware infrastructure necessary to execute large-scale models. Respondent P5 aptly illustrated the issue by questioning the quality of certain published AI models:

\begin{quote}
\emph{``It even happened that model was leaking labels, they were entering the input of the model. And that was not raised as an issue, the paper is still there. You cannot trust the code blindly; otherwise, you repeat the same mistakes.``}
\end{quote}

Despite the existence of Papers with Code, which 6 out of 11 groups of students found helpful in some cases, effectively utilizing benchmarking tools remains a challenge, as pointed out by 2 groups of doctoral students. This difficulty arises from the inherent difficulty of comparing different models, compounded by the vague nature of benchmarks. Consequently, results from automated benchmarking lack reliability and interpretability, as each model or user may define their parameters and occasionally engage in deceptive practices, such as training on testing data or comparing with the weakest model.

\subsection{Other challenges regarding AI research process}
\label{sec:interdisciplinary}
\begin{figure}[t]
\centerline{\includegraphics[height=3.5in]{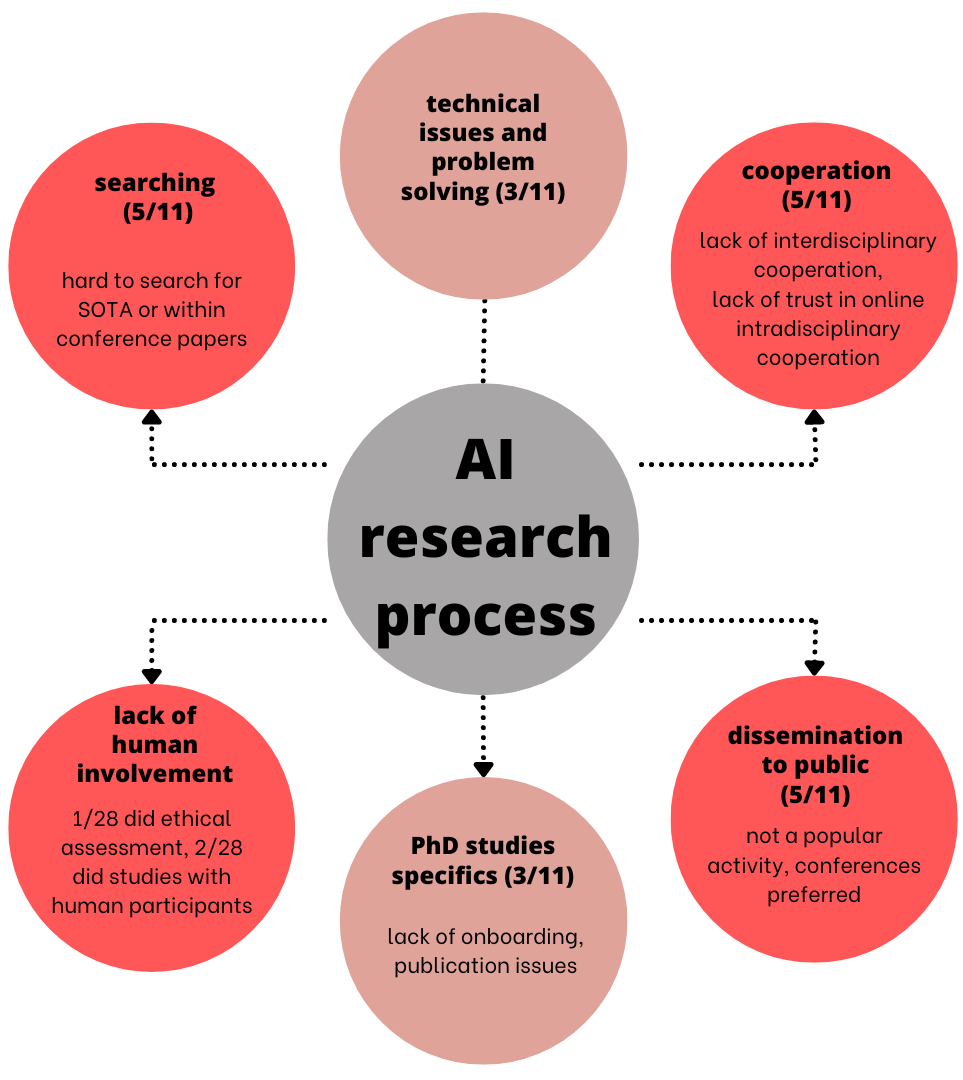}}
\caption{Problematic parts of AI research process as reported by research participants.} \label{research process}
\end{figure}

Our approach allowed us to identify bottlenecks throughout the research process, addressing not only reproducibility challenges but also a spectrum of issues encountered by PhD students.  These include technical challenges and managing expectations related to both AI and doctoral research (both mentioned in 3 groups out of 11), information retrieval, and problems with publication and dissemination (both mentioned in 5 groups out of 11). While not the primary focus of this paper, these issues are significant aspects of the research journey. (Fig.~\ref{research process}). 

These challenges often require collaboration with others. Despite some mistrust in online intra-disciplinary communication, we recognized the need for a scientific interdisciplinary discourse involving experts well-versed in the application domain. Researchers who initiated their research based on their own ideas expressed a desire for increased opportunities to exchange and compare problems while brainstorming with experts from various fields. When such cooperation occurred, it facilitated the advancement of their research. However, researcher P6 encountered difficulties in finding suitable communication channels:

\begin{quote}
\emph{``I store many ideas in my personal document and it is difficult to find someone [outside the field] to communicate with.``}
\end{quote}


We consider another significant concern in AI research to be the lack of human involvement. This concern is closely tied to trustworthiness, primarily regarding accountability requirements. It also intersects with stakeholder participation, as outlined in the ``Diversity, Non-discrimination and Fairness`` requirement within ALTAI (Assessment List for Trustworthy Artificial Intelligence)~\cite{noauthor_assessment_2020}. Young AI researchers seem to overlook ethical assessments and rarely involve human participants in their studies. \emph{De facto}, none of the AI researchers conducted any additional ethical assessments beyond what was required for the ethical sections of the journals. This issue was corroborated by the experience of an AI ethics specialist that has been helping the AI community with these kind of assessments (P8):

\begin{quote}
\emph{``I was surprised by the lack of care of computer scientists about the ethical issues and information about data.``}
\end{quote}

The rationale behind this lies in the constrained understanding of how to conduct ethical assessments, as AI ethics remains ambiguous for many AI scientists. An AI ethics expert confirms the existence of numerous individual frameworks, with many universities lacking standardized templates for consent forms. Within her institution, she addresses the issue through individual discussions, posing questions such as: Are the data ethically collected? Where did it originate and how was it obtained? Who owns or created the data and where is it stored? How long are these data going to be stored?

The minimal engagement of human participants in AI research was a rather unexpected discovery. Only two out of 28 respondents investigated the requirements of their stakeholders or institutional processes, with only one having a carefully planned research design and methodology. However, half of the participants focused their research on areas where involving humans, particularly domain experts, would have been advantageous.

\section{Perspectives and Recommendations}
\label{sec:recommendations}
From our findings, it appears that three main areas of improvements are possible for better AI research: 
\begin{enumerate}
    \item Discoverability and quality of AI resources.
    \item Reproducibility of experiments
    \item Trustworthiness and interdisciplinarity
\end{enumerate}

\subsection{Discoverability and quality of AI resources}
Locating AI resources, such as code, datasets, or models, can be challenging. Our findings highlight that the links between papers and AI resources are often broken once the notification of acceptance sent, making the work of AI researchers more difficult in the long term. 
Although there are specific search engines targeting the discoverability of datasets from multiple sources, such as the Google Dataset Search Engine\footnote{https://datasetsearch.research.google.com/}, these search engines do not extend their purpose to other types of AI resources.
Moreover, the relations among papers and their associated AI resources remains an unexplored domain within these frameworks. Existing platforms such as Papers with Code\footnote{\url{https://paperswithcode.com/}} should, in theory, alleviate the issue, but we observed that this does not seem to be the case.
%
Although such platforms are known quite widely in the AI research community (researchers in 6 groups out of 11 mentioned using it occasionally), they are not always sufficient for these researchers, as there is no quality assurance policy of resources. 
In fact, beyond the discoverability problem, the three categories of AI resources we examined exhibited significant quality issues. We argue that both issues are mostly arising for two reasons, first the lack of appropriate support for the research to perform the work necessary to AI resource curation, and second the lack of incentive and recognition of said work. However, AI researchers could potentially benefit from research data curation services, which encompass tasks such as data cleaning, validation, metadata provision, and consultations. These research services are already integrated into established institutional repositories~\cite{lee_practices_2017}.  

The need for proper documentation of AI resources have been addressed by numerous papers and initiatives, such as the "Datasheets for Datasets"~\cite{gebru_datasheets_2021} or Data Cards~\cite{pushkarna_data_2022}, Dataset cards \footnote{https://huggingface.co/docs/hub/datasets-cards}, and Model Cards for Model Reporting\footnote{https://huggingface.co/docs/hub/model-cards}~\cite{mitchell_model_2019}. Despite increasing outreach and efforts to support researchers in improving the quality of their resources, the problem still persists. As the practices and needs highly depend on the domain, papers summarizing the optimal data curation practices for machine learning emerge, for example in security~\cite{tran_data_2022},  medicine~\cite{diaz_data_2021} or chemistry~\cite{artrith_best_2021}. However, as noted~\cite{rogers_changing_2021}, there is still much work to be done in the realm of data. Therefore, the elaboration of more domain-specific guidelines and best practices are needed. Overcoming interdisciplinary tensions is crucial for progress as well, particularly within AI conferences. 

\begin{recommendation}{collaboration}
    The AI research community needs to foster interdisciplinary collaboration for the governance of AI resources.
\end{recommendation}

Such interdisciplinary collaborations, among others with domain experts, lawyers and ethicists, and data and code curators, must be institutional. Research organizations must provide guidelines and encourage standardized documentation practices. This pertains not only to the quality of AI resources issue, but also the reproducibility of experiments (see Recommendation \ref{institutional-practices}). 

\subsection{Reproducibility of experiments}

Our findings show that the reproducibility of experiments using available AI resources is a huge issue faced by early-career researchers. The causes are a combination of technical and social practice aspects, including the lack of quality in the resources (bugs in code or datasets for instance), but also from discrepancies between papers and associated AI resources, problems in the paper that have not been caught during the peer-review process, or hardware/software that is difficult to access or configure.

To address these issues, the AI research community should get inspired from other fields of research regarding reproducibility practices, such as physics, enforcing stricter policies in journals and conferences. 

\begin{recommendation}{repro-policies}
    The AI research community should embrace reproducibility practices widely and enforce stricter reproducibility policies in journals and conferences.
\end{recommendation}

A small number of conferences and journals in computer science have taken steps to embrace reproducibility initiatives \emph{in practice}. IEEE Transactions on Parallel and Distributed Systems~\footnote{https://www.computer.org/csdl/journal/td/write-for-us/104303} is the first IEEE journal to pilot a reproducibility initiative. Conferences in computer science with reproducibility initiatives include ASPLOS~\footnote{https://www.asplos-conference.org/}, ACM SIGPLAN~\footnote{https://www.sigplan.org/}, ACM SIGMOD~\footnote{https://sigmod.org/}, or SC~\footnote{https://sc24.supercomputing.org/}, and more recently in AI, MLSYS~\footnote{https://mlsys.org/}, or NeurIPS~\footnote{https://neurips.cc/}.

These efforts aim to promote and improve the reproducibility of research by implementing different activities, with reproducibility evaluations being among the most pivotal. Reproducibility evaluations involve dedicated tracks for accepted papers. The authors need to submit a computational resource that includes all the elements to reproduce the experiments in their papers, such as datasets, code, and scripts. 
These initiatives acknowledge that reproducibility encompasses more than mere code and data sharing, and they tackle every aspect that authors need to consider. This includes mechanisms for packaging experiments, such as containers or virtual machines to bypass software dependency issues and streamline portability, installation, and deployment. They also emphasize the importance of curating and documenting datasets, drawing on insights from research on data quality and curation discussed previously. Additionally, they highlight the need to thoroughly document the entire process. As a result of the evaluation, badges can be awarded, which are subsequently recognized by editorials such as ACM or IEEE. These badges recognize a spectrum of reproducibility aspects, including the availability of resources and whether reviewers were able to reproduce the results of the paper. The idea behind this evaluation is to improve the quality of reproducibility in practice.

Furthermore, in addition to create incentives from conference and journals, individual research organization could also contribute to researchers altering their practice. One of our respondents shared with us a noteworthy approach implemented in a robotics-focused laboratory. This lab adhered to mandatory coding guidelines to ensure consistency across projects and employed a dedicated full-time staff member responsible for maintaining code repositories by cleaning and debugging them for all researchers. This practice not only facilitated easier reproducibility but also enhanced collaboration within the lab by enabling efficient reuse of code packages. This is an analogous idea to the aims of these reproducibility initiatives.

\begin{recommendation}{institutional-practices}
    Research institutions and laboratories should set up guidelines and practices for their researchers and provide adequate resources (time and human) to ensure quality of AI resources and reproducibility of experiments.
\end{recommendation}

%

However, even if these reproducibility initiatives were widely adopted, there are still technological challenges that are difficult to address. For instance, the reproducibility of certain experiments with particular hardware dependencies might require very specific hardware. In AI, this can be the case for experiments for constraint programming. 

Currently, editorials do not provide any hardware support yet. Another significant and related challenge is the time required to compile, install, deploy and execute the experiments and, even more, the energy it consumes. First, reviewers face time constraints, and conferences usually impose strict deadlines. Additionally, if authors conduct experiments prior to reviewers conducting their own to verify the reproducibility of experiments, the resulting energy expenditure appears clearly unreasonable.

One solution to these challenges that are not currently addressed could be covered by a cloud federation~\cite{villegas2012cloud, craig2017big}. A cloud federation, as a concept, presents a promising solution for AI researchers to efficiently share AI resources and also computational resources to conduct experiments. A cloud federation also enables to pool together various cloud infrastructures from different providers, then, researchers can access a diverse array of resources tailored to their specific needs, whether it be for data storage, processing power, or specialized hardware like GPUs or TPUs. At the core of facilitating the sharing of AI resources within a federation lies the metadata catalogue, which needs to be designed adopting FAIR principles (i.e. findability, accessibility, interoperability, and reusability). This essential component could enable AI researchers to collaboratively access and utilize a diverse range of resources for their experiments, supporting the navigation among the relationships of datasets, AI models, papers, and experiments. Through the metadata catalogue, researchers can identify and leverage the most suitable resources for their specific research needs, regardless of the cloud provider. This approach not only fosters collaboration but also optimizes resource utilization, ultimately enhancing the efficiency and effectiveness of AI research endeavors.

Finally, the existing tools and services used by reproducibility initiatives for practical reproducibility are not really well-suited for AI and can pose a significant burden. For instance, typically, code is uploaded to a version control system and an open repository like GitHub. To ensure the code's immutability, a tagged version of the code must be paired with Zenodo, which assigns a unique identifier (Digital Object Identifier, DOI) for that specific code version. However, GitHub lacks support for datasets and their versions. Consequently, authors need to turn to \emph{additional} tools and services, such as open dataset repositories, which offer dataset versions and DOIs. Specific tools are required to better match the needs in the context of AI.

\begin{recommendation}{cloud-federation}
    The AI research community should investigate the possibility to create a cloud federation for AI systems and suitable tools for control versioning of AI experiments, including datasets, models and code altogether.
\end{recommendation}

\subsection{Trustworthiness and interdisciplinarity}

The third issue that our research uncovered goes beyond the technical aspects of AI research but more related to interpersonal and institutional relations, and the lack of interdisciplinarity in the field. 

First, we observed that early-career researchers rarely include end-users or human participants in their experiments, despite many of them working on topics that would warrant such practices. The research of PhD students is often very techno-centered: the goal is to produce a working algorithm or system, and the analysis of factors other than performance is often forgotten (for instance user experience, inclusion, ...). Event though methods exist and are considered best practices in other fields, such as Codesign~\cite{trischler2018value}, they are rarely included in the field of AI research. This may be a direct consequence of the second issue that our research highlighted: the lack of interdisciplinary collaboration for early-career researchers. 

Early-career researchers are most often left alone (or with their supervisors) to conduct their research and lack the means to contact or discuss with other researchers, especially outside their institutions or main field of research. Our research also indicates that existing public virtual communities are often not conducive to idea sharing among researchers due to concerns about idea theft.
Similarly to what we recommended for quality and reproducibility (Recommendation \ref{institutional-practices}), institutions have a role to play in encouraging multidisciplinarity. One starting point would be for them to diversify their research staff, for instance with lawyers, ethicists, domain experts, UX or HCI specialists, or information scientists. For institutions in which this diversity already exists (for instance, universities often have several schools that employ researchers in these different domains), they should ensure institutional support for researchers to meet and collaborate on research endeavours.

\begin{recommendation}{diversity}
    Research institutions should support interdisciplinary collaboration by ensuring a large diversity of research staff and providing the means for researchers to collaborate efficiently.
\end{recommendation}

\section{Conclusions and discussion}
\label{sec:conclusion}

%

Good research is made through a collective and multidisciplinary work, where researchers base their work on other researchers' work to advance the collective knowledge. In order to produce good research efficiently, such prior work must be usable by researchers. This means that it must be discoverable, reproducible, and responsible. 

In this paper, we investigated the issues faced by PhD students in the AI field, we discovered that this is not fully the case yet. Despite a growing awareness and initiatives to improve research practices, more efforts need to be made to improve the research process in AI fields. These efforts must be both technological, in developing solutions and platforms to facilitate discoverability and reproducibility, and societal, in encouraging multidisciplinarity and adopting a more trustworthy approach to the research process. To this extent, we proposed five recommendations for the AI research community and institutions to consider in order to improve our research process. 

Our study is limited in the sense that it focused on PhD students and interviewed only a moderate  number of participants (albeit sufficient for this type of research), allowing limited quantification. We have yet to fully ascertain the extent and severity of the uncovered challenges within the community. This stands as a constraint inherent in qualitative research, which typically focuses on uncovering, not quantifying, such challenges. Using quantitative validation methods, like questionnaires, is recommended to validate and quantify these issues, which will be a focus of our future work.

Finally, it is interesting to note that while there has been some amount of work studying the impact of reproducibility (among others related to citation count~\cite{raff2023does,winter2022retrospective}), our work recognized the impact of irreproducibility on both the AI community and its research endeavours. This is still a very unexplored area of research and would warrant more attention.


\section{Research ethical considerations}
This research involved human participants. Research planning, conduct and reporting was consistent with national and international regulatory laws and regulations and aligned with ethical principles and standards for such research. Personal data, such as names and email addresses, were anonymized, and interview data were processed to ensure that sensitive information could not be compromised. All ideas and content are authored by the authors. We utilized AI-based tools (including ChatGPT) to refine the formal aspects of the language.

\ack This work was supported by AI4Europe, a Horizon project funded by European Commission under GA No. 101070000.

\bibliography{bibliography, manuscript}

\end{document}